# Special relativity and the pseudo-Newtonian potential

Marek A. Abramowicz[1], Andrei M. Beloborodov[1,2],
Xingming Chen[1], and Igor V. Igumenshchev[1,3]

[1]Department of Astronomy and Astrophysics, Göteborg University
and Chalmers University of Technology, 412 96 Göteborg, Sweden

[2]Astro Space Center of P. N. Lebedev Institute,
84/32 Profsoyuznaya Street, Moscow, 117810, Russia

[3]Institute of Astronomy, 48 Pyatnitskaya Street, Moscow, 117810, Russia

**Abstract.** A simple re-scaling of velocities calculated from the Paczyński & Wiita (1980) pseudo-Newtonian potential makes them consistent with special relativity and greatly improves the agreement with the exact relativistic calculations. The improvement is relevant in calculations of the Doppler effect, spectra and radiation transfer which all involve effects of special relativity.

## 1. Introduction

The pseudo-Newtonian potential (PN-potential) of Paczyński & Wiita (1980) provides a simple accurate model for black hole hydrodynamics. Under this approximation, one assumes that the gravitational potential of a non-rotating black hole with mass $M$ is

$$\Phi = -\frac{GM}{R - R_G}, \quad R_G = \frac{2GM}{c^2}, \tag{1}$$

and one uses Newtonian theory. Here $R_G$ is the gravitational radius of the black hole. The PN-potential has been used in many papers on black hole hydrodynamics, particularly on accretion disc dynamics (e.g., Okazaki, Kato, & Fukue 1987; Abramowicz et al. 1988). In short, the PN-potential is useful because it is both very simple and very accurate. For example, it predicts the same behaviour of the Keplerian circular orbits of free particles as the exact relativistic theory: around a non-rotating (Schwarzschild) black hole the orbits with $R < 3R_G$ are unstable, and the orbits with $R < 2R_G$ are unbound. Thus, the radius of the marginally stable orbit is located at $R_{ms} = 3R_G$, and the radius of the marginally bound orbit at $R_{mb} = 2R_G$. Accretion discs have their inner edges located somewhere between $R_{ms}$ and $R_{mb}$. Conditions at radii smaller than $R_{mb}$ are not important for the astrophysical properties of accretion discs, but any Newtonian model of the black hole should be accurate for $R > R_{mb}$.

There is a problem with the approach based on the PN-potential. Even though the PN-potential describes accurately the dynamical effects of the curvature of spacetime, it does not include the effects of the special relativity. In particular, it does not guarantee that the velocity of matter should always be less than the speed of light. To see the problem, one may calculate the orbital Keplerian velocity of a test particle, $v$, and the corresponding Lorentz gamma factor, $\gamma = 1/\sqrt{1 - (v/c)^2}$, at the radius of the marginally bound orbit. The correct relativistic value is $v_{BH} = c/\sqrt{2}$ and the corresponding Lorentz gamma factor is $\gamma_{BH} = \sqrt{2}$. However, the values calculated from the PN-potential are $v_{PN} = c$ and $\gamma_{PN} = \infty$. Thus, it is quite obvious that the velocities and the Lorentz gamma factors are not well represented by the PN-potential. Since the Lorentz gamma factors are required in calculations of the Doppler effect, radiative transfer, and spectra, an improvement in calculating the Lorentz gamma factor is helpful in practice.

## 2. Solution

We suggest a re-scaling of the velocities calculated from the PN-potential which provides a simple solution to the problem. The velocity $v_{PN}$ calculated directly from the PN-potential should be replaced by $\tilde{v}_{PN}$ which satisfies

$$v_{PN} = \tilde{v}_{PN}\tilde{\gamma}_{PN}, \quad \tilde{\gamma}_{PN} = \frac{1}{\sqrt{1 - (\tilde{v}_{PN}/c)^2}}. \tag{2}$$

There is a "physical reason" for this re-scaling: if one writes relativistic formulae for the inertial forces in such a way that they resemble their Newtonian counterparts as much as possible, then $v$ in the Newtonian formulae always corresponds to $v\gamma$ in the relativistic formulae (Abramowicz 1992). The re-scaling (2) follows from this general rule. More importantly, however, one may simply check that the re-scaling (2) works amazingly well in practice. In Figure 1a and b we show the velocities and the corresponding gamma factors for Keplerian circular orbits calculated according to the exact relativistic formulae, the PN-potential, and the PN-potential with the re-scaling (2). These velocities are expressed as

$$\beta_{BH} = \frac{v_{BH}}{c} = \frac{1}{\sqrt{2(r-1)}}, \tag{3}$$

$$\beta_{PN} = \frac{v_{PN}}{c} = \sqrt{\frac{r}{2}}\frac{1}{r-1}, \tag{4}$$

$$\tilde{\beta}_{PN} = \frac{\tilde{v}_{PN}}{c} = \sqrt{\frac{r}{2(r-1)^2 + r}}, \tag{5}$$

where $r = R/R_G$. It is seen that, at $r = 2$, $\tilde{\beta}_{PN} = \beta_{BH} = \sqrt{2}/2$. The agreement between the re-scaled velocities and the exact relativistic ones is excellent for $r \geq 2$ as is seen from Figure 1c and d. Specifically, for $r \geq 2$, $\tilde{v}_{PN}/v_{BH} - 1 < 5\%$ and $\tilde{\gamma}_{PN}/\gamma_{BH} - 1 < 2\%$.

The same re-scaling may be used for other Newtonian models of black holes proposed more recently by several authors. All the other models are, however, less accurate and more complicated than that of Paczyński and Wiita, so there is no obvious reason why one should use them (see Artemova, Björnsson, & Novikov 1996)

We conclude that the simple re-scaling of the velocities given by formula (2) makes the pseudo-Newtonian potential of Paczyński and Wiita consistent with the special relativistic effects. Agreement with the exact relativistic theory is better than $\sim 5$ percent. The rescaled velocities calculated from the potential should be used in calculations of all special relativistic effects such as the relativistic Doppler shift, etc.

A.M.B. thanks the Department of Astronomy and Astrophysics, Göteborg University and Chalmers University of Technology for hospitality. He was partially supported by the Russian Foundation for Fundamental Research (project code 95-02-06063).

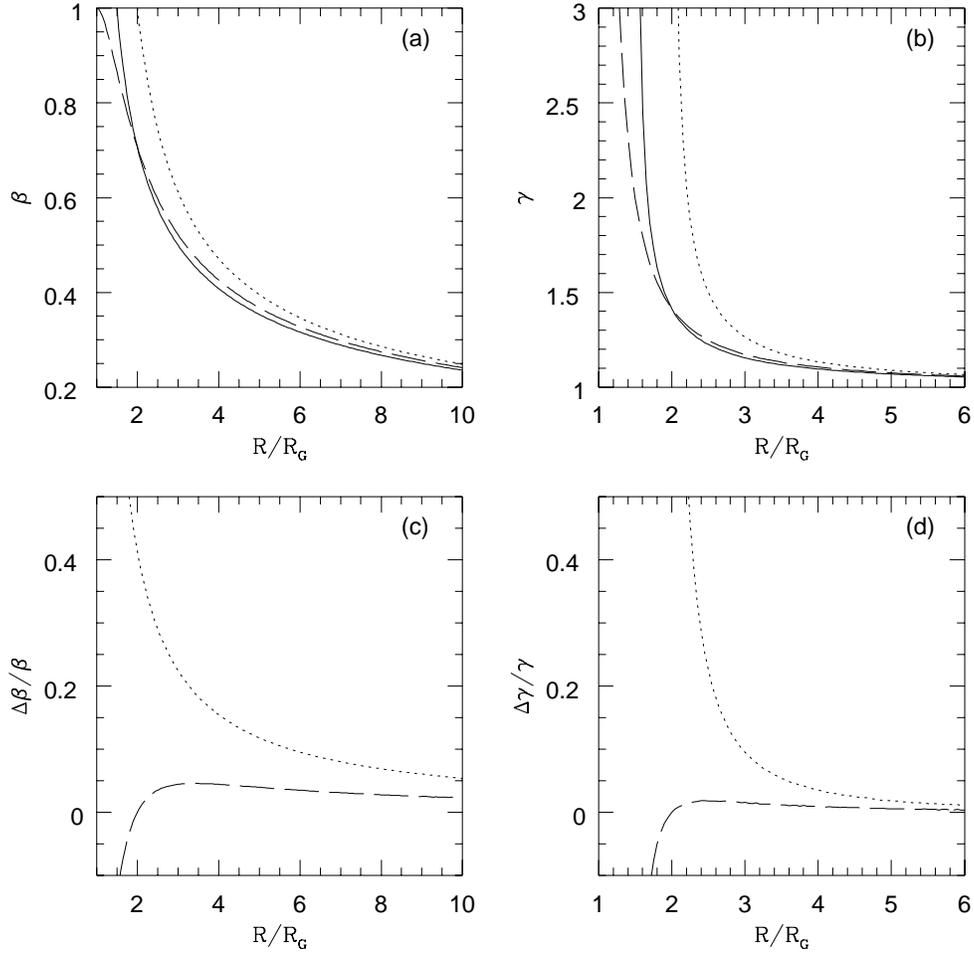

**Figure 1.**— The Keplerian orbital velocities (panel a) and the corresponding Lorentz gamma factors (panel b) calculated according to the exact relativistic formulae (solid line), the PN-potential (dotted line), and the PN-potential with the re-scaling (dashed line). The relative errors of the velocities (panel c) and the Lorentz gamma factors (panel d) calculated from the PN-potential with the re-scaling (dashed line) and without the re-scaling (dotted line). Note that at $R = 2R_G$, the PN-potential with the re-scaling gives the same result as that of the exact relativistic theory, and the agreement between these two is excellent for $R \geq 2R_G$.